\documentclass[11pt,aas2pp4]{aastex}

\slugcomment{Submitted to the Astrophysical Journal Supplements}

\shortauthors{Roberts, Romani, and Kawai}
\shorttitle{The ASCA Catalog of GeV Sources}

\begin{document}

\title{The ASCA Catalog of Potential X-Ray Counterparts of GeV Sources}

\author{Mallory S.E. Roberts\altaffilmark{1} and Roger W. Romani}
\affil{Department of Physics, Stanford University,
    Stanford, CA 94305-4060}
\email{mallory@astro.stanford.edu, rwr@astro.stanford.edu}

\and

\author{Nobuyuki Kawai}
\affil{Cosmic Radiation Laboratory, The Institute of Physical and
Chemical Research (RIKEN), 2-1 Hirosawa, Wako, Saitama 351-0198, Japan}

\altaffiltext{1}{Current address: McGill University, Physics Department, 
3600 University St., Montreal, Quebec, H3A 2T8}

\begin{abstract}
We present a catalog of 2-10 keV ASCA Gas Imaging Spectrometer images of fields 
containing bright sources of GeV emission. The images cover $\sim 85\%$ of the
95\% confidence position contour for 28 of the 30 sources with $> 1$ GeV fluxes
above $4\times 10^{-8}$ ph ${\rm cm}^{-2} {\rm s}^{-1}$. We find an 
excess of hard X-ray sources with $F_{2-10 keV} \ga 10^{-12}$ ergs 
${\rm cm}^{-2} {\rm s}^{-1}$ positionally coincident with unidentified sources
of GeV emission. We comment on radio-loud and radio-quiet pulsar candidates,
as well as several SNR and massive binaries as possible sources of 
the GeV emission.
We also find evidence for a class of variable $\gamma-$ray sources associated 
with extended regions of hard X-ray emission, and propose that for these
sources, a significant percentage of the $\gamma-$ray emission is generated in 
synchrotron nebulae surrounding fast, young pulsars.
\end{abstract}

\keywords{gamma rays: observations --- pulsars: general --- stars: Wolf-Rayet --- stars: X-ray --- supernova remnants}

\section{Introduction}
 
Despite more than 20 years of study, the majority of bright,
high energy $\gamma-$ray sources along the Galactic plane have yet 
to be identified with lower energy counterparts. With the life of 
the EGRET instrument on CGRO now over, nearly 300 sources 
have been detected and catalogued above 100 MeV.  At high Galactic 
latitudes ($b\ga 10^{\circ}$),
approximately 70-90 of these sources have been identified with a sub-class of 
radio-loud active galactic nuclei (AGN) called blazars \citep{h99}. 
At low latitudes ($b\la 10^{\circ}$), 7-8 sources have been identified as
young, rapidly spinning pulsars. In addition, there is a source associated
with the LMC, and a detection of a solar flare. The rest of the sources
have no firm identifications, but can be split into at least three components. 
First, there is an isotropic component which is likely to consist largely of 
blazars with slightly lower radio fluxes than the ones already identified
(since most of the brightest sources have already been identified). There
is an excess of sources at mid-latitudes ($10^{\circ}\la b \la 30^{\circ}$), 
making up a second component, which has been suggested to be
associated with the Gould Belt \citep{gr98}, with some indication of
an additional component associated 
with the Galactic halo. Finally, there is a population
which is highly concentrated along the Galactic plane, and is our main
concern here. 

Spectral studies \citep{me96} suggest a need for additional source classes. 
In particular, there are many sources with steeper spectra (photon index
$\Gamma\ga 2.2$) between 100 
MeV and 1GeV than is expected for either a pulsar or a blazar. In 
addition, variability studies \citep{m96,t99} indicate an excess
of variable sources at low Galactic latitudes. Since pulsars are not thought
to be variable, this would suggest an additional class of  Galactic
sources. However, for any individual source to represent a new source 
class, it would first have to be demonstrated that it is not a 
blazar, which are highly variable. 

The main obstacle to identification has been the large
positional uncertainties of the sources. Low counting rates combined 
with a broad, energy-dependent point-spread function (PSF) produce
typical 95\% error contour sizes of $\sim0.5^{\circ}-1^{\circ}$. Additional
problems are encountered in the Galactic plane, where sources are
often confused and the diffuse background is strong. Background models
are based on radio maps which are not sensitive to small scale structure,
causing additional uncertainty. However,  the Galactic background spectrum
steepens at energies above 1GeV and the EGRET PSF narrows with
increasing energy. Therefore, sources with significant counts above 1 GeV are
less prone to the systematic positional errors associated with sources
in the Galactic plane {\it if} the improved resolution of the high energy photons
is included in the analysis. Unfortunately, the standard analysis of EGRET data
uses a binned likelihood approach, where all the photons are assumed to have
the same PSF derived by averaging over the energy range using an assumed
spectrum, so the additional information to be gained from the
photon energy is lost. In the third EGRET catalog \citep{h99}, 
this issue was partially
addressed by the generation of liklihood test statistic maps for positional
determination in three energy ranges: $> 100$ MeV, $300-1000$
MeV, and $> 1000$ MeV, and the map that gave the best positional determination
was used (in the case of some of the stronger sources, the $300-1000$ MeV
and $> 1000$ MeV maps were combined). However, the source candidate list used
was generated only from the $> 100$ MeV maps.

The purpose of this work is to begin the process of systematically 
identifying potential low energy counterparts of the unidentified
$\gamma-$ray sources (for a more thorough
treatment of one  source, GeV J1417-6100, see Roberts and Romani 1998; 
Roberts et al. 1999; Roberts 2000). 
To start, we note a few observational facts about the known
source classes. First, they all emit non-thermal 
X-rays. The blazars are moderately bright ($F_x \simeq
10^{-12}-10^{-11}$ ergs ${\rm cm^{-2} s^{-1}}$ in the
2-10 keV band) X-ray point sources with power law spectra. They are also bright 
($\sim 1-10$ Jy at 5 GHz), compact, radio sources. Their
emission is highly variable at most wavelengths. 

Pulsars can be faint($\sim10^{-13}$ ergs ${\rm cm^{-1} s^{-1}}$) 
or bright ($\sim10^{-8}$  ergs ${\rm cm^{-1} s^{-1}}$ in the case of 
the Crab) sources of non-thermal X-rays.
The magnetospheric component of this flux is strongly pulsed. Additional
X-ray and/or radio emission may come from a surrounding wind
nebula.  There may also be a thermal X-ray component from the pulsar
surface or a surrounding SNR.
Pulsars tend to have very steady emission at most wavelengths
(save for the pulsations). 

There are several other proposed sources of GeV emission, most notably supernova remnants accelerating particles through shocks with either 
the interstellar medium or a nearby 
molecuar cloud (cf. Pohl and Esposito 1998). The particle spectrum predicted
by these models could produce significant synchrotron emission in
the X-ray region, depending on the age of the SNR and on the magnetic
field and density of the SNR and surrounding region.
Non-thermal X-ray emission has been observed in the SNR IC443
\citep{k97}, suggesting it as a potential source of this class. However, this class of 
objects would not be expected to produce variable emission unless the site
of the shock was highly localized, and 
the ability to separate this emission from potential emission from
an embedded pulsar is generally not possible without better $\gamma-$ ray
resolution. Other suggested sources of
particle acceleration through shocks, such as interacting winds of massive
binary stars \citep{eu93}, or
the wind off of a supermassive star \citep{vf82}, would also be likely to
produce hard X-ray emission at some level. The more localized 
nature of the emission would make variable emission feasible.

Another proposed source class of $\gamma$-ray emission is isolated
rotating black holes accreting from the interstellar medium \citep{p99a,an99}.
This class is attractive since it is expected to be variable on 
all time scales. However, the emission mechanisms are not well developed,
and predictions of the ratio of X-ray to $\gamma$-ray emission are hard
to make.

Several studies \citep{yr97,kc96} have shown the Galactic
unidentified EGRET sources to be associated with regions of star formation.
In particular, the distribution of low latitude EGRET sources is strongly 
correlated with SNR and OB associations, which supports the above
proposals for new source classes. However, young pulsars are also
found near SNR and OB associations, so the statistical studies in and of 
themselves do not require the existence of a new source class.  
From an observational viewpoint, since the sources are near regions
of star formation, they are likely to be even more heavily obscured by
gas than is typical for sources in the Galactic plane. This reinforces the
need for X-ray observations above 2 keV. 

In order to search for (or rule out) potential counterparts, we
have obtained observations of the brightest GeV sources with the ASCA satellite
\citep{tih94}.
We will primarily focus on 
the Gas Imaging Spectrometer instruments (GIS2 and GIS3)\citep{o96}, 
whose good
sensitivity out to 10 keV and large field of view (44') are well suited for
finding potential counterparts. 
 
\section{Source Selection and $\gamma-$Ray Analysis}

We base our source selection on the GeV source catalog of Lamb and Macomb (1997, hereafter LM).  Through a combination of archival and new pointings, we
have obtained ASCA GIS images of every field containing a source with  a $\gamma-$ray flux above 1 GeV greater than 
$5\times 10^{-8} {\rm ph}\,{\rm cm}^{-2}\, {\rm s}^{-1}$, and all but 1 of the unidentified
sources (GeV J1814-1228) and 1 blazar (PKS 1622-297) with fluxes above $4\times 10^{-8} {\rm ph}\,{\rm cm}^{-2}\,{\rm s}^{-1}$.  
This flux limit corresponds to the ``bright" sources of LM. While most of these sources are
also listed in the third EGRET (3EG)
catalog \citep{h99},  several are not, even though they are quite significant
at GeV energies. The 3EG catalog is based on sources found to be significant 
in a binned likelihood analysis of all photons above 100 MeV detected by 
the EGRET telescope, while LM used only those photons above 1GeV. 
Of the unidentified sources in this study, the GIS fields cover $\sim85\%$
of the $95\%$ error contours, and  virtually all of the $68\%$ contours. 
The coverage of each individual source is listed in Table~\ref{OBStab}.

This focus on sources with significant GeV emission will tend to
select out pulsar candidates, which are expected to have relatively flat
$\gamma-$ray emission out to several GeV (cf. Romani 1996).
It also limits the number of potential unidentified
blazars in the survey. Of the 30 sources in LM that make our flux cut, only 
6 are at Galactic latitudes $|b|>11^{\circ}$, 5 of which are known blazars. Our
expectation of the number of additional blazars in the survey is $\la 1$.

The LM catalog was derived solely from the $\ge1$ GeV 
count maps. However, nearby sources which
are significant in $\ge100$ MeV maps but not in the GeV maps can 
still produce enough GeV photons to bias a position
towards the softer source if it is not included in the fit. In addition,
the elliptical fit to the $95\%$ error contour is often not a good
approximation. Therefore, we feel the LM localizations are not
adequate in many cases. 

Most of the GeV sources are in the 3rd EGRET catalog, and for 
many of these, we have used the likelihood test statistic (TS) maps available from the on-line catalog 
to generate the positional contours on our images. 
Most of the maps used are based on 1GeV and above photons. 
In four cases (GeV J0008+7304, GeV J0241+6102, GeV J1837-0610, and GeV 
J1856+0115) we used the $>300$MeV positional contours which were 
fully consistent with the GeV contours, but better constrained.
Several sources are not in the
3EG catalog, mis-identified, have only low-energy maps, or are near sources not in the 3EG catalog (such as
in the Cygnus region). For these we generated new TS maps, using the
{\it like} program of John Mattox and Joe Esposito (jaelike5.49, Mattox et al.
1996), on maps of
1GeV and above photons. We included in the fits all nearby sources in the 
3$\sigma$ list (courtesy R. Hartman) which was used to create the 3rd EGRET 
source catalog. 
The fluxes derived from these fits are systematically lower than in LM, 
since a portion of the GeV photons may be assigned to the softer 
sources which were not included in the LM fits. Where we have refit
the data, we use the fluxes derived from those fits. Otherwise, we use
the LM values. These are listed in Table~\ref{MULtab}. As a consistency check, we have 
also examined the TS maps of \citet{ml00}, which only included
sources which had a $>3\sigma$ significance in the GeV maps for the positional
fits. In general, the fits were similar except for noticable changes in
contour shape of some of the sources with soft sources nearby, as expected. 
The 3EG catalog lists photon spectral indices, and for those sources with
firm 3EG ids (see Table~\ref{OBStab}), we have included those values in 
Table~\ref{MULtab}. 

All of the 3EG sources were searched for variability by Tompkins (1999), 
using the $\tau$ statistic, which is the standard deviation 
of the flux divided by 
the average flux, and hence a measure of how variable a source is (as
opposed to the more usual $\chi^2$ test, which measures  how
inconsistent a source's flux is with being constant). The individual 
flux measurements used to determine $\tau$ were derived from 3 parameter fits
to the entire unbinned likelihood distribution of the flux above 100 MeV.
A flux was derived for each viewing period  ($\sim$ two weeks) where the 
source was within $25^{\circ}$ of the pointing center.  
Therefore,
the timescale of the variability probed is $\sim 1$ month --
$\sim 2$ years. The pulsars all tend 
to have $\tau\sim 0.1$, consistent within systematic uncertainty with 0,
while blazars tend to $\tau \ga 1$. Extreme caution should be used when 
interpreting the variability of sources in crowded regions,
since the likelihood analysis may occasionally misassign photons, resulting in 
time bins with  anomolously high or low fluxes. In addition, variability in a 
nearby source may result in an apparent variability of the source of 
interest. With these warnings, the $\tau$ values, where available, are 
also listed in Table~\ref{MULtab}.

\section{X-ray Observations}

Table~\ref{OBStab} gives the observing parameters of all the fields in our 
survey. 
The observations were taken over several years, with differing exposure
times. Some of the archival observations had the GIS in modes with limited
spectral or spatial resolution and the imaging and spectral analysis
was adapted accordingly. In a few cases, the SIS data were also used. 
Ten of the images were obtained specifically for this campaign. 
In these cases, the
pointings were based solely on the GeV positions. Positional and spectral 
information on X-ray sources found in these fields are given in 
Table~\ref{NEWfit}.
Some of the pointings were based on 
LM97 values, and subsequent positional fits have resulted in a significantly
shifted error contour. 

Stellar sources with soft spectra are common in the plane, and
can be quite bright below 2 keV, but are generally undetectable above 2 keV.
In addition, the sources we are searching for are likely to be 
behind fairly high hydrogen column depths, and are expected to
have hard, synchrotron spectra. To make these sources more apparent,
we produce 2-10 keV images.

For this work, we used the standard rev 2 processed data. This only
aspects the inner $44^{\prime}$ of the image, and region filters are 
applied to remove the calibration source and the outer edges where there
is poor aspecting.  Processing was mainly done using the $FTOOLS$ and
$XSPEC$ packages available from the HEASARC, while image compositing
and display 
were done using the $MIRIAD$ and $KARMA$ packages available from the ATNF
\citep{sk98,g95}. 

\subsection{X-Ray Imaging}

As mentioned above, our primary instruments for this work are the two
Gas Imaging Spectrometers on board the ASCA satellite. The effective area
of these instruments is strongly dependent on both energy and position
on the detector. If an exposure correction is not applied, broad
diffuse artifacts will occur in the image from the general radial 
dependence of the instrument, with sharp features due to the 
grid support structure.  
The specific pattern is energy dependent and the artifacts tend
to be stronger at higher energies. It is
therefore necessary to create an exposure map based on the spectrum
of the X-ray field. In principle, this could, and would, vary across the
field from different sources at different positions. However, our primary
purpose is the detection of sources above the relatively smooth Galactic
background.  We therefore chose a relatively blank field at a typical
Galactic position to derive our field spectrum (an observation of $\alpha-
Centauri$, which is at $l=315.733 , b=-0.681$ and has virtually no emission 
above 2keV, worked well
for this purpose). Note that this is not ideal for fields dominated
by a large, extended source with a spectrum significantly different from
the background, such as is the case of fields with a bright, thermal, SNR.
The ftool ascaeffmap was initially used to derive an efficiency
map from this spectra based on ground calibrations of the detectors. This was
then used in the ascaexpo tool created
by Eric Gotthelf to build up an exposure map in sky coordinates by using
the individual observations attitude file to calculate how much time each
detector pixel would contribute to each sky pixel.

When the resulting exposure map was divided into the image map, it was found to
overcorrect
the image in the center. There are two reasons for this. The first is that
the GIS detectors are sensitive to the particle background, which tends to
be brightest towards the edges of the detector, in opposition to the
exposure effects. We therefore used the night earth calibration
observations with identical screening as our images, normalized their
exposure, and ran it through ascaexpo to create sky particle maps with
exposures identicle to the images. We used these to subtract off the
particle contribution of the images, before applying exposure corrections.
The second cause is the scattering of X-rays onto the
detector from regions outside the field of view, which causes an apparent
enhancement in the background towards the edges of the detectors. Since
we are interested in having as flat of a background as possible in order to
be confident that any structure we see is real, we created an efficiency
fudge map by subtracting the particle background from the deep blank sky images
provided by the ASCA GOF, dividing by our model efficiency map derived 
from the ascaeffmap tool, and then 
highly smoothing it to show only the broad effects
of the scattered light. This fudge map was then multiplied by the original
efficiency map
which gave us a final detector efficiency map whose fine structure was based on
the Galactic plane spectrum, but which has broad structure to suppress
scattered light from an assumed flat background.

One effect of the exposure correction on low count rate fields is the
enhancement of noise at the edges of the fields where the exposure drops 
off suddenly. Pixels containing photons will have apparent fluxes greatly
exaggerated when divided by a relatively low exposure. This is
especially noticeable if there is a source just outside the
field of view, increasing the amount of light scattered onto the field edge.
To minimize these edge effects, we trim the exposure maps
so as to blank the image once the exposure drops below a 
threshold. Even so, edge artifacts are still apparent in many of the images, 
so apparent emission near a field edge, even in composited images, should
be viewed with skepticism.

The steps used to make an image are then as follows: 1.  create photon, 
particle background, and exposure maps; 2. subtract the background from
the photon map; 3.  sum the gis2 and gis3  photon and exposure maps, 
and, if there is more than 1 field, composite the photon 
maps and exposure maps separately; 4. trim the exposure maps; 5.
divide the summed photon maps by the summed exposure maps to create 
intensity maps; 6. smooth the intensity maps with a $\sigma=1.7$ pixel
Gaussian, which roughly corresponds to the $50^{\prime \prime}$ core width 
of the PSF. In addition, the coordinate offsets derived by \citet{got00}
to correct for the temperature dependent deviation in the attitude solution
were applied.
Source positions for point-like sources were derived from centroiding.
Positional errors are dominated by systematic uncertainties, estimated
to be about $24^{\prime \prime}$ for sources within the central $20^{\prime}$
of the detector. Sources toward the edges of the detector have additional 
uncertainties of $\sim0.5^{\prime}-1.0^{\prime}$ as they approach the 
detector edge. We note these in Table~\ref{NEWfit}. The faintest sources may
also have additional systematic errors due to source confusion.
Positions of extended sources are
the center of the extraction regions used for spectral fitting, 
or of the peak emission if there is a significant peak.

\begin{figure}[ht!!]
\plotone{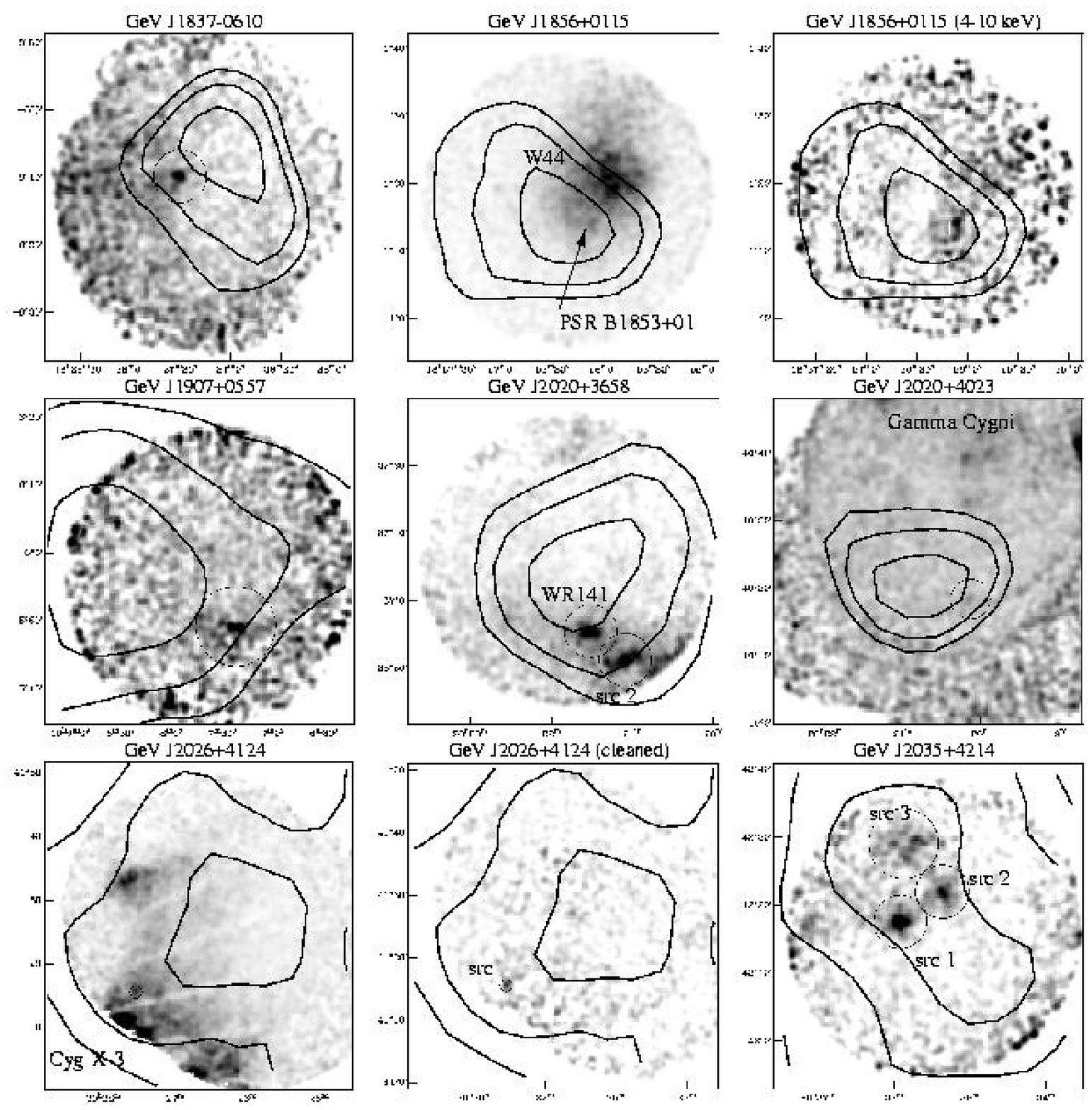}
R.A. (2000) and Dec. (2000). The contours are the 
\caption{\label{IMAGE} The ASCA GIS image catalog.  Coordinates are in
R.A. (2000) and Dec. (2000). The contours are the
68\%, 95\%, and 99\% confidence positions of the GeV source. Unless noted
otherwise, the energy range is 2-10 keV. Circles represent spectral extraction
regions. The cross shaped appearance of bright point sources (or
compact extended sources like the Crab) is due
to the ASCA PSF.}
\end{figure}

The images are shown in Figure~\ref{IMAGE}. A few sources 
are dominated by thermal emission from a bright supernova remnant, 
in which cases we also show 4-10 keV
images. Scattered light from very bright
sources can contaminate fields over 1 degree away from the source,
appearing as diffuse emission with linear features. Two (possibly three) of 
the images are affected by this.

\pagebreak

\begin{figure}[ht!!]
\onecolumn
\plotone{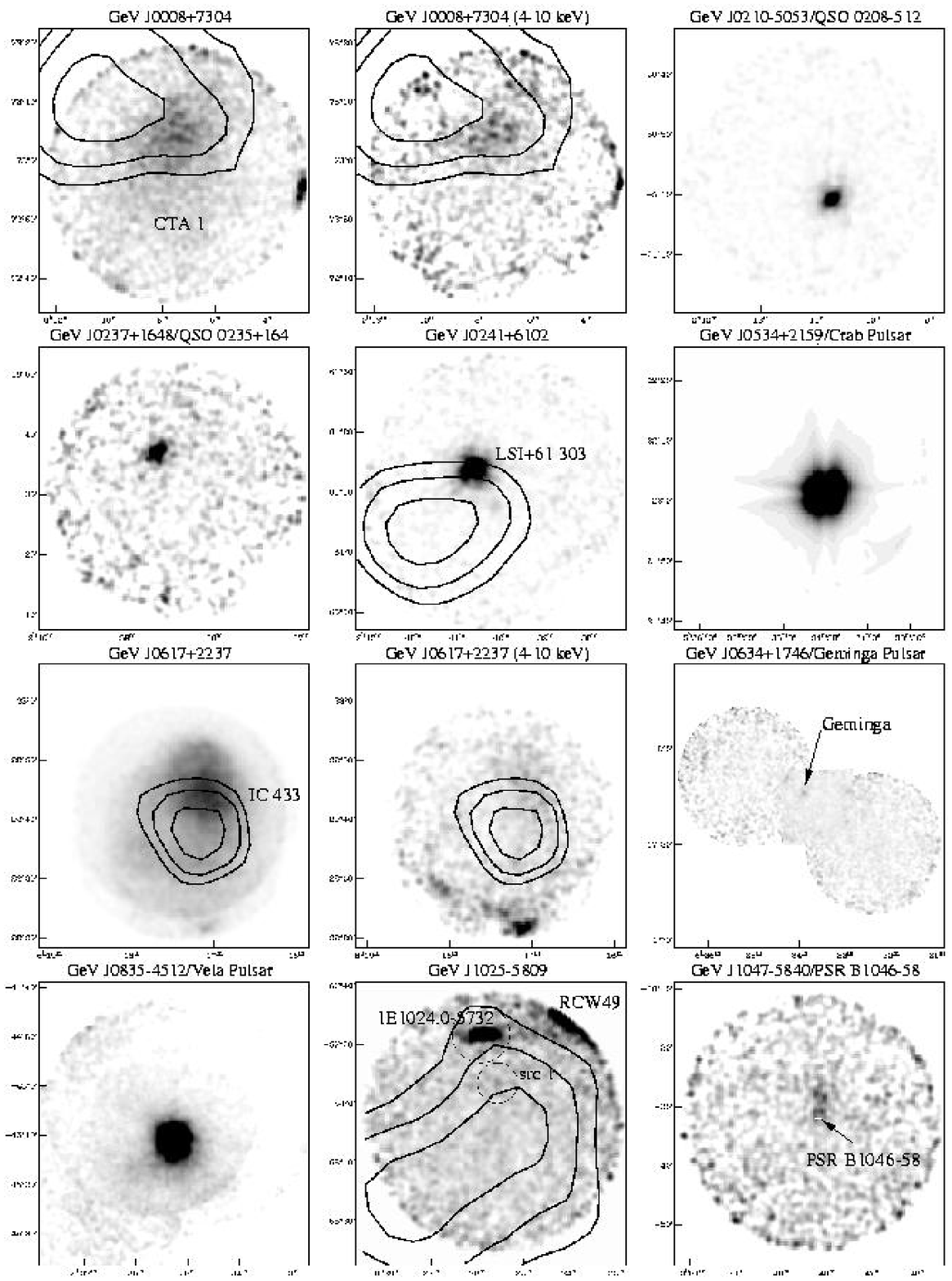}
\end{figure}
\clearpage
\begin{figure}[ht!!]
\plotone{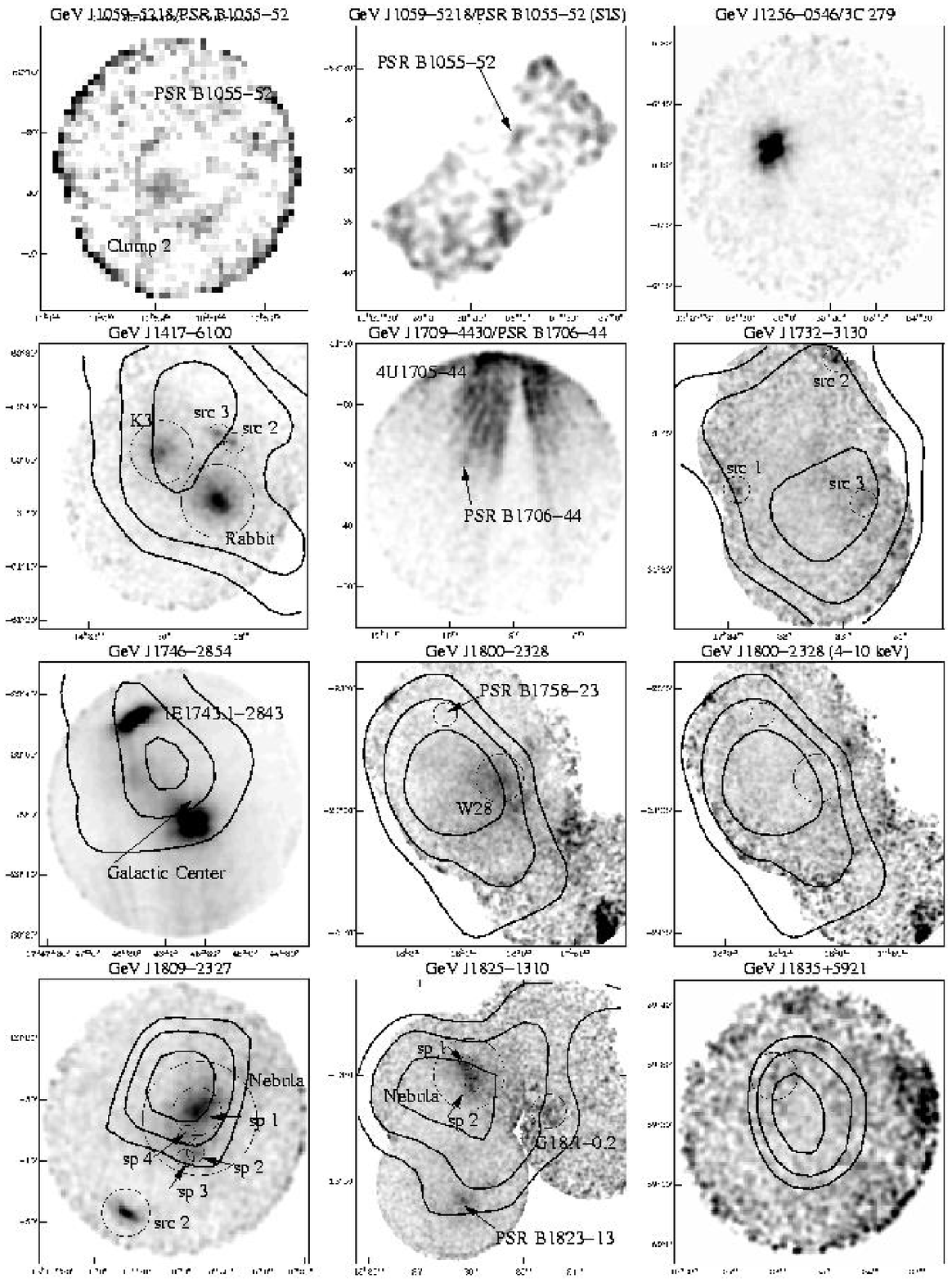}
\end{figure}
\clearpage

\twocolumn

Some of the archival data of fields containing pulsars were obtained in modes with 
low spatial resolution. In these cases, SIS images are shown.
The fits images and efficiency maps are available on the web at
http://astro.stanford.edu/$\sim$mallory/survey.html.

\subsection{Spectral and Flux Measurements}

Many of the sources discovered are faint, and we want to 
ensure that detector artifacts are not mistaken for real sources. Therefore, 
extra care has to be taken in spectral measurements.
Spectra were extracted from the brightest sources discovered in the images
as indicated in the figures.
Since most of the sources are in the Galactic plane, background regions 
were chosen, where possible, from within the same field. The 1999 version
of the night
earth observation files were used to subtract the particle background from 
both the source and background spectra. Since the sensitivity is not constant 
over the field of view, a correction needs to be made in the background
spectra. To do this, we determined the average 2-10 keV detector efficiency
in the source and background extraction regions using the efficiency map 
used in the exposure corrections described above. We then scaled the 
background spectra (after the night earth spectra was subtracted) by the ratio
of the background to source efficiency.  
To test how well this corrected for the dependence on detector region of the
background spectra, we took four spectra from different regions of the
detectors in the field of GeV J1025-5809 and applied our corrections. 
We then fit both the faint ($\sim 3\times 10^{-13} {\rm ergs}\,{\rm cm}^{-2}\,
{\rm s}^{-1}$) and moderately bright ($\sim 3\times 10^{-12} {\rm ergs}\,
{\rm cm}^{-2}\,{\rm s}^{-1}$) sources in the field using each of the 
backgrounds in turn. All of the fits measured fluxes within $1 \sigma$ of
each other, and the derived spectra were all consistent. Thus, we believe
our method is adequate for deriving consistent flux measurements for the
new sources in this survey independent of where on the detector the 
background was extracted.

The spectra from the two GIS detectors were then 
combined and fit with an absorbed power law model using the XSPEC
package. The results are listed in Table~\ref{NEWfit}. 
The best fit parameters with
$90\%$ multi-parameter error
intervals are listed in the table. The flux measurements $F_X$ are of the 
absorbed 2-10 keV flux and have $1 \sigma$ 
errors derived from single parameter fits of the normalization. 
If the source was too faint for a meaningful fit, an absorption
column of $1\times 10^{22} {\rm cm}^{-2}$ and a spectral index of 2.0
was assumed in measuring the flux. 

Table~\ref{MULtab} lists all of the ``bright" sources from \citet{lm97} ($F_{GeV} > 4 
\times 10^{-8}$ ph ${\rm cm}^{-2} {\rm s}^{-1}$), with the names and
X-ray fluxes of some of the more interesting counterpart candidates, 
and the $\gamma-$ray 
fluxes, spectral indices, and the $\tau$ variability statistic where available.
In the cases where there is a bright, thermal SNR in the field, 
we first fit the spectrum using a thermal plasma
model, and then added a power law component with a spectral index of 2.0 
to see if the fit improved
and to provide upper limits on any non-thermal component.
Only the non-thermal flux limit is listed in Table~\ref{MULtab}. 
In cases where there was no obvious source, the same fitting process
was used to provide $1 \sigma$ upper limits on the flux. In some cases of
previously known objects, we quote values from the literature. 

In order to compare the different sources, we calculated an X-ray to
$\gamma-$ray energy ``spectral index" $\alpha_{X\gamma}=
1+log(F_{\gamma}/A_X)/6$ where $F_{\gamma}$ is the photon flux above 
1GeV, which
corresponds to the flux density at 1GeV if a photon index of 2.0 is
assumed, and $A_X$ is the X-ray power law normalization in 
photons ${\rm GeV}^{-1}\,{\rm cm}^{-2}$ 
, corresponding
to the unabsorbed flux density at 1keV. 
This value is also listed in Table~\ref{MULtab}. 

\section{Individual Sources}

We have split the sources into three categories. The first, the focus of this study, consists of sources which had previously 
not been associated with known sources of high energy emission (such as 
supernova remnants, pulsars, or X-ray binaries). These fields were selected based upon $\gamma-$ ray
source positions.  In order to estimate the significance of the X-ray sources
found in these fields, we use the log N-log S relationship of
Sugizaki (1999) derived from the ASCA Galactic Plane Survey (GPS) covering
the area of $|l| < 40^{\circ}$ and $|b| < 0.4^{\circ}$. For each listed
source found within or near the $95\%$ positional contour, we list in
Table~\ref{NEWfit} the number
of sources of that flux or greater expected to be found within the $95\%$ 
contour. Note that many of the sources are at larger Galactic latitudes and
longitudes than the GPS, and so this should be a fairly conservative
estimate of the chance of random association.  

The second category consists of fields which were pointed
at some previously  known lower energy source, whose position is
coincident with a GeV error ellipse. The third category consists of 
GeV sources which have positive identifications with a low energy
counterpart (i.e. a pulsar or a blazar). 

\subsection{GeV Selected Sources}

{\it GeV J1025-5809}
This source is near the bright radio source RCW49, an HII region with 
colliding wind bubbles from massive stars at the center 
\citep{wu97} seen as
a significant X-ray source at the NW edge of the field. The
peak marked as src1 may be associated with this extended
emission. The point source
just outside the $95\%$ confidence contour is the 
probable Wolf-Rayet+O star binary 1E1024.0-5732/Wack 2134 \citep{re99}.
There is much extended radio emission in the region associated with RCW49,
including a blob suggestively shaped like a bow shock, with no apparent 
X-ray emission.
 
{\it GeV J1417-6100}
This source has been the object of recent extensive studies in radio and
X-rays \citep{rr98,r99,cb99,rr00}.
The two extended sources (marked as K3 and Rabbit in the figure) 
are associated with non-thermal 
radio emission in the wings of the Kookaburra complex
\citep{r99}. The more southerly one is coincident with
the Rabbit nebula, whose
radio spectral and polarization properties 
suggest a pulsar wind nebula (PWN) origin. However, the Parkes multi-beam 
survey has recently discovered the high $\dot E$ pulsar 
PSR J1420-6048 consistent with the
northern extended source \citep{dam00} which also
has radio properties consistent with a PWN (K3 of Roberts et al. 1999). 
A point radio source near the peak of the K3 source 
is the likely pulsar \citep{rr00}. 
AX J1418.2-6047 (src2 of Roberts and Romani 1998, and so marked in figure) has no
associated radio emission, and may be variable. The $\gamma-$ray source 
also appears to be
moderately variable, and therefore this last source could be a candidate
for an isolated black hole. However, given the 
presence of at least one, if not two, potential
pulsar counterparts, we do not favor this latter possibility.

{\it GeV J1732-3130}
This is listed as another name for 3EG J1734-3232 in the third EGRET
catalog, but the rather large $>100$  MeV $95\%$ error contour
only slightly overlaps the LM GeV ellipse, making their identification as
a single source unlikely.  Our new TS map assumes an additional source
at the best fit 3EG position, but this has little effect on the final result.
The large contour required 2 ASCA pointings for adequate coverage. 
The X-ray image shows weak, extended emission within the $68\%$ contour, however
this may be scattered light from the X-ray binary X1724-308, just over 1 degree
away. We measured a peak within this area (src3), and two sources near the
edges of the FOV (src1,src2). We could find no other
X-ray images of this field.

{\it GeV J1809-2327}
This image shows an X-ray complex with possible
point sources (sp2 and sp3) in the southern portion coincident 
with massive young stars in the Sharpless 32 HII region. 
There is also a coincident 60 $\mu$ source seen in the IRAS survey
with a peak at 18h 09m 58s, -23d 41m 14s (near sp3).
However, the $\gamma-$ray contours favor the northern part of the complex, 
which has a harder spectrum. This extended X-ray emission is 
surrounded by molecular gas in the Lynds 227 dark nebula, and has been
suggested to be a synchrotron nebula maintaining pressure equilibrium
with the cloud by means of a pulsar wind (see Oka et al. 1999 for details). 
This source also seems to be moderately variable in $\gamma-$rays. 
The bright point source to the south (src2) is near a weak, radio point source.
Such X-ray/radio sources are frequently associated with Seyfert galaxies,
which are common background sources in hard X-rays \citep{g90}.

{\it GeV J1825-1310}
This source is near the young pulsar PSR B1823-13. However,
our GeV source position is not consistent with it at the $95\%$ confidence 
level.  The
new image, based on the GeV position, reveals a previously unknown
extended X-ray source with a spectrum suggestive of a pulsar wind nebula.
Near the conjunction of our field with the archival GPS fields is 
some apparently thermal diffuse X-ray emission near a non-thermal radio source in the
Sharpless 53 HII cluster, which may be a supernova remnant (G18.1-0.2, 
marked as such in figure, also source
F of Kassim et al. 1989). This latter source is also consistent with the soft 
$\gamma-$ray source ($\Gamma = 2.69 \pm 0.19$) 3EG J1823-1314. 

{\it GeV J1835+5921}
This source is the only unidentified GeV source that made our flux cut at
high Galactic latitude ($b\sim 25^{\circ}$). Due to its hard spectrum,
small error contour,  and unique position resulting in low absorption, 
it has been the subject of an extensive observing campaign with
ROSAT and ASCA, as well as radio and optical studies \citep{m00}.
Only a few, very faint point sources are in the field. 
\citet{m00} suggest one variable soft X-ray source observed by ROSAT
as a potential counterpart. 
Strong upper limits
on the optical flux suggest a neutron star identification. 
However, the relative X-ray to $\gamma-$ray flux is remarkably low
assuming the X-rays are due to
thermal emission from a neutron star surface.  
This may indicate an aging isolated pulsar with only weak magnetospheric 
X-ray emission. In our image, made from the archival data,
the most significant peak consistent
with the GeV source is at 18h36m9.2s, +59d28m09s, which is a marginal
source with a 2-10 keV flux of only $\sim 10^{-13}{\rm ergs}\,{\rm cm}^{-2}
\,{\rm s}^{-1}$. This is not one of the sources listed by \citet{m00}.
However, we use this as the X-ray flux in Figure~\ref{FXfg} and 
Figure~\ref{AXtau}.

{\it GeV J1837-0610}
This field contains a single point source in a small, well constrained 
error contour. The field was observed twice, with slightly different
pointings, and the point source was seen at the same sky coordinates
in both. 
The Parkes multi-beam survey has dicovered a fast pulsar 
within the contour \citep{dam00}
which is not consistent with the X-ray source. 

{\it GeV J1907+0557}
This source is listed as 3EG J1903+0550 in the third EGRET
catalog, even though the center of the LM ellipse was $\sim 1^{\circ}$
away with hardly any overlap with the very large $95\%$ contour. It
seems likely that 3EG J1903+0550 is associated with the SNR G40.5-0.5,
in which case it has no association with GeV J1907+0557. Our image
was based on the LM position, and contains two weak pointlike 
sources which may be bright spots in a single,
extended, source, or fluctuations from a low count rate. Our flux 
measurement treats them as one source.

{\it GeV J2020+3658}
There were two previous ASCA pointings attempting to see this source
based on the position of the second EGRET catalog, which was to the
northeast of this field. In the third EGRET catalog, the source was split into
two sources, the southern one, 3EG J2021+3716, being consistent with the 
GeV source position (NOT 3EG J2016+3657, as listed in the 3EG catalog).
Interesting extended emission can be seen in our image, with 2 
embedded sources. The one more centered in the GeV
contour (src1) is coincident with WR141, 
a WN6+O type Wolf-Rayet binary system with a
21.6 day period \citep{l96}. Several lines are seen in the spectra, and
thermal plasma model fits to src1 result in $kT\sim5$ keV. Src2 is somewhat 
harder, with a smoother spectrum, suggesting it is more likely to be
non-thermal.  
The extended emission appears to be moderately absorbed (nH$\sim 10^{22}
{\rm cm}^{-2}$) with a moderately steep spectrum ($\Gamma \sim 2.5$), 
although accurate flux and spectral measurements are difficult due
to the two point sources dominating the flux in the region. This
diffuse emission is also seen in an archival Einstein IPC image. 

{\it GeV J2026+4124}
The ASCA image of this region suffers badly from scattered X-rays
coming from Cyg X-3, even though it is over $1^{\circ}$ away. However,
Cyg X-3 is variable, and by creating maps from times at low count rates
from which we subtract scaled maps from high count-rate times, we
discover at least one source in the GeV contour (``cleaned" image). 
This source is seen in both pointings that make up the composite image. 

{\it GeV J2035+4214}
This field contains three interesting sources, two point-like and one
extended. Src1 is coincident with a bright ($\sim 1$Jy
at 1420 MHz), 
steep spectrum double-lobed radio source \citep{c92}, unusual 
in that for one of the lobes the rotation measure has the 
opposite sign from the other lobe, and from that of 
most of the sources in the region.
The second source (src2) is embedded
in the radio-bright (44 Jy at 408 MHz, Normandeau, Joncas, and Green 1992) 
dense molecular cloud DR17, which has a
positive velocity measure (+10 km/s, Piepenbrink and Wendker 1988), 
indicating it is probably in the 
Great Cygnus Rift ($d \la 1$ kpc). 
The third 
X-ray source (src3) is extended and has a somewhat softer, possibly thermal spectrum. 

\subsection{Sources Coincident With Previously Known Sources}

{\it GeV J0008+7304; CTA 1}
The X-ray nebula seen in this
image, contained within the SNR CTA 1, has been interpreted as 
synchrotron emission powered by the
wind of a fast pulsar, proposed to be the ROSAT point
source RXJ0007.0+7302 found near the center \citep{s97}. 
Limits on the optical and radio flux have lead Brazier et al. (1998)
to propose this to be a radio-quiet $\gamma-$ray pulsar. 
The X-ray and $\gamma-$ray spectra
are consistent with this hypothesis. The source may 
be moderately variable however, suggesting that at least some of the 
$\gamma-$ray emission is not from a pulsar magnetosphere. 

{\it GeV J0241+6102; LSI+61 303}
This source is nearly coincident with the curious X-ray binary LSI+61 303.
The source is variable in $\gamma-$rays, and although this has not been
correlated with any of the known timescales of LSI+61 303 \citep{kn97},
several authors have suggested potential mechanisms of $\gamma-$ray
production (see Strickman et al. 1998 and references therein; Punsly 1999a).
However, it should be noted that the source is now barely excluded at the $95\%$ confidence level
in the 3rd EGRET catalog. The faint peak just to the southwest of 
LSI+61 303 has been identified as a stellar source \citep{lhy97} which is unlikely to
be associated with the $\gamma-$ray source. No other source is apparent
in the field.

{\it GeV J0617+2237; IC443}
Thermal X-ray emission from the SNR IC443 dominates the 2-10 keV
image of this field. Several authors have pointed to this
source as an example of how cosmic ray production can produce GeV
emission from a SNR, either alone or interacting with a molecular cloud
(eg. Hnatyk and Petruk 1998).
Indeed, Keohane et al. (1997) discovered 2 spatially localized
regions of non-thermal emission that can be seen here in the 4-10 keV
image at the southern and south-eastern edges of the field. 
Both of these regions are well outside the updated GeV
error contour, and we find little evidence for a non-thermal component at
the $\gamma-$ray position, although the bright thermal emission could
easily obscure a faint source.

{\it GeV J1746-2854; Galactic Center}
Emission in the Galactic center region is very complex at all wavelengths, 
and the GeV emission is no exception. The X-ray image shows several 
potential counterparts, and the shape of the GeV error contour suggests
it may result from several, confused sources. The northern source
within the 95\% GeV contour is the X-ray binary 1E 1743.1-2843, which
we use in the table,
is connected by a ridge of emission to the source or sources near the
Galactic center which are just outside the 95\% contour.
For detailed discussions
of potential $\gamma-$ray sources in this region, see \citet{mh98}.

{\it GeV J1800-2328; W28/PSR B1758-23}
The error contour of this source encompasses the thermal emission
from the SNR W28 and the young pulsar PSR B1758-23. We see no 
evidence of hard emission at the pulsar position, and no strong non-thermal
emission in the GeV error contour, although, like in IC443, a weak source could
easily be obscured by the bright thermal emission. 
The 3rd EGRET catalog
$\gamma-$ray spectrum is somewhat softer than most of the sources in this
study, and we note that the catalog position, 
based on the $> 100$ MeV maps, extends further to the south, 
being consistent with the non-thermal emission seen in the 
4-10 keV image at the southwestern edge of the field.
It is possible that the softer $\gamma-$rays come from this region, while
the GeV emission comes from PSR B1758-23 or a radio-quiet pulsar 
hidden in the center of W28.

{\it GeV J1856+0115; W44/PSR B1853+01}
\citet{h96} have discovered an extended non-thermal source associated
with PSR B1853+01, seen here in the 4-10 keV image. \citet{f96} have
discovered an associated PWN. However, the $\gamma-$ray emission 
appears to be variable, and no pulsations have been detected in the
EGRET data. 

{\it GeV J2020+4023; $\gamma$ Cygni}
This source, located within the $\gamma$ Cygni supernova remnant, has
a very well constrained error ellipse. Despite extensive observations, 
there is no obvious hard X-ray component associated
with the GeV source. Our extraction region provides an upper limit to
any hard X-ray counterpart in this field.

\subsection{Identified Sources}

As noted above, the identified sources fall into two classes: blazars and
pulsars. There are 5 blazars listed by LM among their ``bright" sources,
however only 2 have GeV fluxes above 
$5\times 10^{-8} {\rm cm}^{-2} {\rm s}^{-1}$. 
The blazars in this list observed by ASCA all appear as moderately bright
point sources with power law spectra. In the single dish PMN radio survey,
these sources all have bright, point-like 5 GHz emission. 
The $\alpha_{X\gamma}$ values of these sources are fairly narrowly 
confined between -0.65 and -0.75. 

The pulsars are much more difficult to characterize. At one extreme is the 
Crab, with a very bright X-ray nebula, $\alpha_{X\gamma}=-1.28$, and a
bright radio pulsar wind nebula (PWN).
At the other extreme is Geminga, with very weak X-ray emission, 
$\alpha_{X\gamma}=-0.32$, and no radio emission. 
PSR B1046-58 and PSR B1055-52 have moderately
weak X-ray emission, with possibly associated extended emission, and no
evidence of a radio PWN. PSR B1706-44 is similar, but may have a weak
radio PWN. Although difficult to see in the image here which strongly
suffers from scattered X-rays from the nearby X-ray binary 4U1705-44,
there is some evidence for a compact X-ray nebula around the 
pulsar \citep{f98}. Vela has moderately strong extended X-ray and 
radio emission. The pulsars all have low variability, consistent with 
none when systematics are considered, except for Vela. The apparent
small variability of Vela is generally assumed to be due to the presence of
nearby artifactual sources \citep{t99}.
  
\section{Discussion}

\begin{figure}[h!]
\plotone{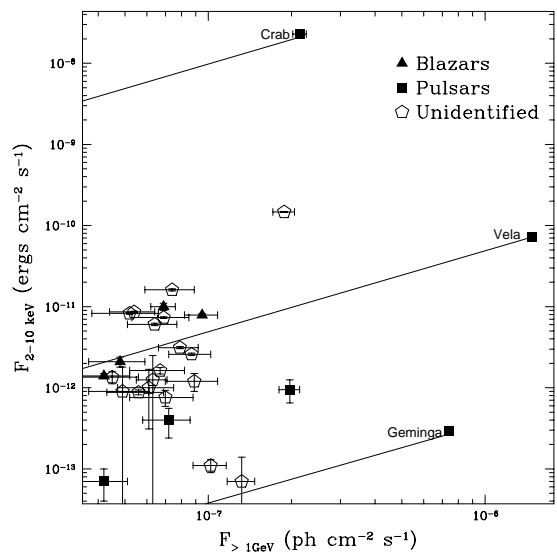}
\caption{\label{FXfg} GeV flux vs. 2-10 keV flux of the 
brightest source consistent with the GeV position, and therefore 
an upper limit on the counterpart flux. Lines of constant
$F_X/F_{\gamma}$ are drawn through the Crab, Vela, and Geminga pulsars.}
\end{figure}

As can be inferred from the values listed in Table~\ref{NEWfit}, 
there is clearly an
excess of sources with fluxes of a few $\times 10^{-12}$ ergs 
${\rm cm^{-2} s^{-1}}$ when compared to the GPS logN-logS, 
implying many of these are associated with the GeV sources. In some fields,
however, multiple candidates make individual IDs impossible. In 
Figure~\ref{FXfg}, we
plot the 2-10 keV X-ray flux of the brightest (and therefore statistically
the source with the least likelihood of a chance coincidence) 
potential counterpart in each 
field versus the GeV flux. Since we are plotting the brightest source, 
these points can be considered upper limits on the ``true" X-ray
counterpart.  For reference, we include the known pulsars and 
blazars on the plot, with lines of constant flux ratio drawn through the Crab,
Vela, and Geminga pulsars. These are representative of pulsars with 
characteristic ages of roughly $10^3$, $10^4$, and $10^5$ years. Note there
are no Crab-like sources, the closest being the source near the Galactic
center. A few sources have X-ray fluxes or upper limits which result in 
ratios consistent with Geminga. Then there is a clump of sources with Vela-like
ratios, which include the blazars. However, upper limits from
5 GHz radio surveys rule out a blazar ID in almost all cases for the
unidentified sources.  

\begin{figure}[h!]
\plotone{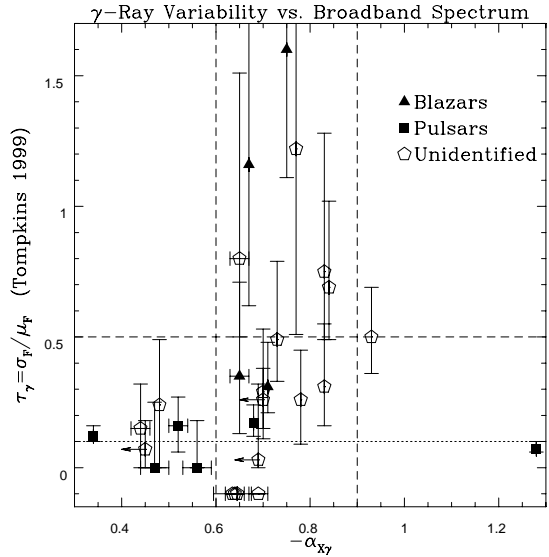}
\caption{\label{AXtau} Broadband spectral index vs. $\gamma$-ray
variability above 100 MeV. The dotted line is the systematic variability.
The four sources plotted with negative values have not been measured for
variability. Vertical dashed lines are region boundaries for
X-ray faint, X-ray moderate, and X-ray bright sources. The horizontal dashed 
line separates the low variability and high variability zones. The
values of $-\alpha_{X\gamma}$ are from the brightest likely counterpart,
and therefore can be viewed as upper limits. Except where plotted, 
errors on $-\alpha_{X\gamma}$ are negligible ($\la 0.01$).
}
\end{figure}

In Figure~\ref{AXtau}, we plot these same sources' energy index 
$-\alpha_{X\gamma}$ {\it vs.} the $\tau$ $\gamma-$ray variability measure. 
Again, since we are using the brightest source, the value of $-\alpha_{X\gamma}$
in this plot for each source can be considered an upper limit on the true
value, so if the plotted candidate X-ray source is not the GeV source,
the source position on the plot would generally move to the left.
The dotted line represents the systematic variability value $\tau = 0.1$. 
Sources not in the 3EG catalog, thus lacking a
measured variability, are plotted on the bottom at $\tau=-0.1$.   
The dashed lines split the plot between sources with high and low variability, and
what we will call (from left to right) X-ray faint, X-ray moderate, 
and X-ray bright. 
In the latter category, there is only the Crab and possibly the Galactic center 
source.
We immediately note that all of the identified pulsars are low variability, with
the four isolated pulsars being X-ray faint, Vela being X-ray moderate, and the
Crab being X-ray bright.
Three other sources are in the low-variability, X-ray faint category: GeV 
J1835+5921, GeV J1837-0610, and GeV J2020+4023. GeV J1025-5809
is in this category if the Wolf-Rayet star 1E1024.0-5732, outside the $95\%$
positional contour, is not the GeV source. In addition, if PSR B1758-23 is the
counterpart for GeV J1800-2338, it would also fall within this region. 
All of these sources in this category are good candidates for isolated pulsars,
about half of which are coincident with radio pulsars.
 
In the X-ray moderate category, we have four SNR. Two of them, IC443
and W28, only have upper limits on non-thermal X-ray emission associated
with the GeV emission. In W44, the extended non-thermal emission is associated
with a pulsar wind nebula around PSR B1853+01, and in CTA 1, there is a
proposed association of the extended non-thermal emission with the 
radio-quiet pulsar candidate RXJ0007.0+7302 \citep{s97,b98}. 
Therefore, we do not see any indication of GeV emission generated by SNR
shocking with the interstellar medium or molecular clouds, 
although we cannot rule
out this scenario. However, it should be noted that in some cases 
error contours
produced from $>100$ MeV maps are consistent with regions of apparently
shocked emission, and that several of the softer sources in the
3rd EGRET catalog are coincident with young SNR. It therefore seems 
likely that SNR produce high-energy $\gamma-$rays, but with a steeper
spectrum than pulsars and blazars.

Three of the X-ray moderate candidates are binary systems. Two, 1E1024.0-5732
and WR 141, are WN5-6+O
star systems whose hard X-ray emission is almost certainly due to shocks from
colliding winds. With only $\sim 20$ known WN+O systems, and a viable 
emission mechanism, the coincidence of these two similar systems with 
GeV sources is suggestive. Neither source is strongly variable at 
$\gamma-$ray energies, although in the case of WR 141 the natural time scale
of the binary period is shorter than the variability 
analysis is sensitive to.  
 However, both of these sources cannot be considered strong
candidates since one is outside the 95\% confidence contour and the other has
another equally bright, and hence equally viable, X-ray point source consistent
with the GeV position. LSI+61 303 is another candidate, but in this 
case the particle acceleration mechanism is less clear. If the companion to the B star
is a neutron star, the $\gamma-$rays could be produced through an accretion
process, colliding winds, or even standard pulsation mechanisms. 
Its moderate variability would suggest that some of the emission could
be due to interactions of the neutron-star with its environment.  

The Vela pulsar is in the X-ray moderate, low-variability group. 
When the binary sources are eliminated, only four sources are in this
region of the plot, although there are an additional four potential members
of this group which were not tested for variability. Of these, two sources
are weak upper-limits on non-thermal flux hidden in thermal SNR, and 
are therefore, if pulsars, likely to be X-ray faint. The best candidates
for Vela-like radio-quiet pulsars are therefore 
RXJ0007.0+7302 in CTA 1, and src2
in GeV J2020+3658. Src2 in GeV J2035+4214 and the double source in 
GeV J1907+097 are reasonable candidates among the sources with unknown 
variability. Predictions of the majority of unidentified GeV sources
being radio-quiet, Vela-type pulsars by some outer-gap models of
pulsar emission (eg. Yadigaroglu and Romani 1997) 
are therefore not well supported by the X-ray data.     

The most intriguing sources are the X-ray moderate, 
high variability unidentified
sources. Although consistent with the blazars in the survey, none have
bright, compact radio sources coincident with the X-ray source. 
What is most remarkable is that all four sources in this category
seem to contain extended, synchrotron nebulae. Two of these, GeV J1856+0115
and GeV J1417-6100, have nebulae 
associated with the high $\dot E$ pulsars PSR B1853+01
and the Kookaburra Pulsar PSR J1420-6048. 
The other two, GeV J1809-2327 and GeV J1825-1310,
are in star forming regions providing potential pulsar  birth-sites, with
the latter having as a secondary source candidate PSR B1823-13. GeV J0008+7304,
the source
associated with CTA 1 which
also has a synchrotron X-ray spectrum, shows indications of $\gamma-$ray
variability and could also be included in this group. 

The most likely explanation for these sources are wind nebulae
around young pulsars. The variability would indicate that a substantial
fraction of the $\gamma-$ray flux is synchrotron/Compton emission generated by
particles in the pulsar wind. Depending on the local magnetic field strength,
the synchrotron cooling timescale could be on the order of a few months, 
similar to the variability time scale. De Jager et al. (1996) have suggested a
similar explanation for a possible variation seen in the Crab at 70-150 Mev. 
If this is the case, the synchrotron spectrum would mostly
dominate the lower energies, and the $\gamma-$ray pulse fraction should increase
with energy until the pulsar emission cuts off at several GeV. Oka et al. (1999)
propose that in the case of GeV J1809-2327, the 
$\gamma-$ray emission is bremsstrahlung photons from a pulsar wind colliding
with target baryons in the molecular cloud Lynds 227. In this case, the variability 
could be due to instabilities in the interaction layer.

\section{Conclusions}

We have presented a nearly complete 2-10 keV X-ray image catalog of potential
counterparts to the brightest sources of GeV emission. In the images of
the unidentified
sources, we find an excess of X-ray sources with 
$F_{2-10keV} \ga 10^{-12}$ ergs ${\rm cm^{-2} s^{-1}}$ within
the 95\% error contours. 
Among these, we find several candidates for isolated, radio-quiet pulsars. 
We find no evidence for GeV emission from supernova remnant shell
shocks, although we cannot yet exclude this possibility. We also find 4-5 
extended hard X-ray sources coincident with 
variable $\gamma-$ray sources, representing a potential new class of GeV
sources, plausibly associated with pulsar wind nebulae. The sources in this
catalog are ready targets for the new generation
of X-ray satellites. In particular, Chandra could map the fine structure 
of the candidate PWN, compare their morphologies with known PWN, and search
for the pulsar location. XMM, with its high throughput, could obtain excellent
spectra and search for X-ray pulsations. Detailed radio imaging, spectroscopy,
and polarimetry could prove very enlightening for many of these sources. 
Ultimately, it will take further $\gamma-$ray measurements to determine
which of these sources are true low-energy counterparts. The improved
spatial resolution and high-energy sensitivity of the proposed GLAST mission
should easily make unambiguous identifications of most of these sources. 

\acknowledgments

We would like to thank E. Gotthelf for invaluable aid with the image analysis,
V. Kaspi for sharing preliminary results,
and P. Nolan, B. Tompkins, and B. Jones for useful discussions on interpreting
EGRET data. We would also like to acknowledge R. Hartman for providing us
with the 3$\sigma$ list of 3rd EGRET catalog sources, and D. Macomb for 
TS maps of his GeV analysis. Roger W. Romani is a Cottrell scholar of the
Research Corporation.
This work made use of several HEASARC on-line resources, including Skyview and
W3Browse, and
was supported in part by NASA grants NAG 5-3333 and NAGW-4562.

\clearpage

\begin{deluxetable}{lccccc}
\tabletypesize{\scriptsize}
\tablecaption{ASCA Observation Parameters \label{OBStab}}
\tablewidth{0pt}
\tablehead{
\colhead{Source} & \colhead{Coverage \tablenotemark{a}} & \colhead{Date} &
\colhead{Exposure} & \colhead{3EG ID} & \colhead{Notes} \\
& & MJD & ksec & &
} 
\startdata
GeV J0008+7304 & 0.68 & 50107.35 & 44 & 3EG J0010+7309 & CTA 1?\\
GeV J0210-5053 & ID\tablenotemark{b} & 49924.84& 12 & 3EG J0210-5055 & QSO 0208-512\\
GeV J0237+1648 & ID & 49387.82 & 12 & 3EG J0237+1635 & QSO 0235+164\\
GeV J0241+6102 & 0.99 & 49386.09 & 17 & 3EG J0241+6103 & LSI+61 303?\\
&& 49392.34 & 19 & & \\
GeV J0534+2159 & ID  & 51071.43 & 43 & 3EG J0534+2200 & Crab Pulsar\\
GeV J0617+2237 & 1.00 & 49091.34 & 19 & 3EG J0617+2238 & IC 443 SNR?\\
&& 49092.11 & 23 & & \\
GeV J0634+1746 & ID & 49439.90 & 74 & 3EG J0633+1751 & Geminga Pulsar; SIS\\
&& 50521.95 & 38 & & OFFSET \\
&& 50523.12 & 18 & & OFFSET \\
GeV J0835-4512 & ID & 49119.09 & 10 & 3EG J0834-4511 & Vela Pulsar\\
GeV J1025-5809 & 0.78 & 50834.95 & 37 & 3EG J1027-5817 & \\
GeV J1047-5840 & ID & 49379.14 & 18 & 3EG J1048-5840 & PSR B1046-58\\
GeV J1059-5218 & ID & 49735.14 & 27 & 3EG J1058-5234 & PSR B1055-52; SIS \\
&& 49735.14 & 37 && GIS lo-res	\\
GeV J1256-0546 & ID & 49159.94 & 28 & 3EG J1255-0549 & 3C 279\\
GeV J1417-6100 & 0.96 & 50316.32 & 45 & 3EG J1420-6038 & Kookaburra Nebula?\\
&& 51222.67 & 59 &&\\
GeV J1709-4430 & ID  & 49605.31 & 17 & 3EG J1710-4439 & PSR B1706-44\\
GeV J1732-3130 & 0.98 & 51254.85 & 21 && mis ID as 3EG J1734-3232\\
&& 51254.26 & 20 &&\\
GeV J1746-2854 & 1.00 & 49610.93 & 82 & 3EG J1746-2851 & Gal. Center region\\
GeV J1800-2328 & 1.00 & 49445.08& 28 & 3EG J1800-2338 & W28/PSR B1758-23?\\
&& 49445.82& 21 && \\
&& 49448.95& 10 && \\
&& 50173.30& 12 && \\
&& 50173.56 & 10 && \\
GeV J1809-2327 & 1.00 & 50525.75& 71 & 3EG J1809-2328 & \\
GeV J1825-1310 & 0.90 & 51255.25 & 37 &3EG J1826-1302 & mis ID 3EG J1823-1314\\
&& 50548.28 & 10 && GPS\tablenotemark{c} \\
&& 50547.88 & 8 && GPS \\
GeV J1835+5921 & 1.00 & 50923.84 & 69 & 3EG J1835+5918 & \\
GeV J1837-0610 & 1.00 & 50905.31  & 20 & 3EG J1837-0606 &\\
&& 51104.87 & 18 & &\\
GeV J1856+0115 & 1.00 & 49464.93 & 14 & 3EG J1856+0114 & W44/PSR B1853+01?\\
GeV J1907+0557 & 0.74 & 51096.08 & 23  && mis ID 3EG J1903+0550\\
GeV J2020+3658 & 1.00 & 51299.67 & 43  & 3EG J2021+3716 & mis ID 3EG J2016+3657\\
GeV J2020+4023 & 1.00 & 50219.45 & 39 & 3EG J2020+4017 & $\gamma$Cyg SNR?\\
&& 50582.71 & 59 && \\
&& 50584.50 & 16 && \\
&& 50584.83& 12 &&  \\
GeV J2026+4124 &0.72 & 51339.69 & 21 & &\\
&& 51340.28 & 22 &&\\
GeV J2035+4214 & 0.74 & 50948.81 & 37 & \\

\enddata

 
\tablenotetext{a}{95\% contour covered by all pointings combined}
\tablenotetext{b}{ID indicates $\gamma-$ray source has been identified with a 
low energy counterpart.}
\tablenotetext{c}{GPS indicates field is from the ASCA Galactic Plane Survey}

\end{deluxetable}

\clearpage

\begin{deluxetable}{lccccccc}
\tablecaption{Multiwavelength Comparison \label{MULtab}}
\tablewidth{0pt}
\tabletypesize{\scriptsize}
\tablehead{
\colhead{$\gamma-$ray source}&\colhead{Candidate}&
\colhead{$F_{2-10keV}$\tablenotemark{a}}&\colhead{$D$\tablenotemark{b}}&
\colhead{$F_{\gamma > 1GeV}$\tablenotemark{c}}&\colhead{$\Gamma_{\gamma}$}&
\colhead{$\tau$}&
\colhead{$-\alpha_{X\gamma}$\tablenotemark{d}} \\		
&&$10^{-12}$ ergs ${\rm cm}^{-2}{\rm s}^{-1}$&
kpc&$10^{-8}$ ph ${\rm cm}^{-2}{\rm s}^{-1}$&&&}
\startdata
\multicolumn{8}{c}{GeV Selected}\\
GeV J1025-5809&AX J1025.6-5757&$0.34\pm0.07$&5&$8.7\pm1.5$&
$1.94\pm0.09$&$0.26^{+.19}_{-.17}$&0.45(2)\\
&AX J1025.9-5749&$2.59\pm0.09$&3&$8.7\pm1.5$&
$1.94\pm0.09$&$0.26^{+.19}_{-.17}$&0.78 \\
GeV J1417-6100&AX J1420.1-6049&$4.82\pm0.17$&
1.5&$6.9\pm1.6$&$2.02\pm0.14$&$1.22^{+6.21}_{-.71}$&0.68\\	
GeV J1732-3130&AX J1732.2-3044&$1.35\pm0.23$&-&$4.5\pm1.1$&-&-&0.69(2) \\
GeV J1809-2327&AX J1809.8-2333&$3.65\pm0.08$&1.9&$5.4\pm1.0$&$2.06\pm0.08$&$0.69^{+.33}_{-.20}$&0.76\\
GeV J1814-1228&-&-&2&$4.6\pm1.1$&-&-&-\\
GeV J1825-1310&AX J1826.1-1300&$8.27\pm0.24$&4.1&$5.2\pm1.4$&
$2.00\pm0.11$&$0.75^{+.53}_{-.26}$&0.83\\
GeV J1835+5921&AX J1836.2-5928&$0.11\pm0.02$&-&$10.2\pm1.4$&$1.69\pm0.07$&$0.15^{+.17}_{-.15}$&$0.44(2)$\\
GeV J1837-0610&AX J1837.5-0610&$1.63\pm0.15$&9&$6.7\pm1.5$&
$1.82\pm0.14$&$0.24^{+.25}_{-.24}$&0.48\\
GeV J1907+0557&AX J1907.1+0549&$0.76\pm0.17$&2.1&$7.0\pm1.8$&
-&-&0.64(2)\\
GeV J2020+3658&AX J2021.1+3651&$3.83\pm0.13$&5&$7.9\pm1.3$&$1.86\pm0.10$&$0.29^{+.24}_{-.18}$&
0.69\\
GeV J2026+4124&AX J2027.6+4116&$1.00\pm0.69$&1.6&$6.1\pm1.4$&-&-&0.63(4)\\
GeV J2035+4214&AX J2035.4+4222&$0.88\pm0.06$&1&$5.6\pm1.3$&-&-&0.64\\
\multicolumn{8}{c}{Previously Known Candidate}\\
GeV J0008+7304&CTA 1&$16.1\tablenotemark{e}$&2.1&$7.4\pm1.5$&
$1.85\pm0.10$&$0.31^{+.24}_{-.15}$&0.83\\
GeV J0241+6102&LSI+61 303&$6.05\tablenotemark{f}$&2.4\tablenotemark{g}&$6.4\pm1.3$&
$2.21\pm0.07$&$0.49^{+.30}_{-.16}$&0.73\\
GeV J0617+2237&IC443&$< 2.5\tablenotemark{h}$&1&$6.3\pm0.8$&$2.01\pm0.06$&
$0.26^{+.12}_{-.11}$&$< 0.70$\\
GeV J1746-2854&Gal. Center&$150$&1&$18.8\pm1.7$&
$1.70\pm0.07$&$0.50^{+.19}_{-.14}$&$0.93$\\
GeV J1800-2328&PSR B1758-23&$< 0.09\tablenotemark{h}$&3&$4.9\pm1.2$&
$2.10\pm0.10$&$0.03^{+.29}_{-.03}$&$<0.48$\\
&W28&$< 1.8\tablenotemark{h}$&3&$4.9\pm1.2$&
$2.10\pm0.10$&$0.03^{+.29}_{-.03}$&$<0.69$\\
GeV J1856+0115&PSR B1853+01/W44&$1.2\pm0.3\tablenotemark{i}$&3.3&$8.9\pm1.9$&
$1.93\pm0.10$&$0.80^{+.71}_{-.30}$&0.65(2)\\
GeV J2020+4023&$\gamma$ Cygni&$<0.14$&0.7&$13.2\pm1.5$&
$2.08\pm0.04$&$0.07^{+.11}_{-.07}$&$<0.45$\\
\multicolumn{8}{c}{Identified Sources}\\
GeV J0210-5053&QSO 0208-512&$7.9\pm0.3\tablenotemark{j}$&-&$9.5\pm1.3$&
$1.99\pm0.05$&$0.31^{+.17}_{-.10}$&0.71\\
GeV J0237+1648&QSO 0235+164&$1.4\pm0.1\tablenotemark{j}$&-&$4.2\pm1.0$&
$1.85\pm0.12$&$1.16^{+2.72}_{-.54}$&0.67\\
GeV J0534+2159&Crab&$23000\tablenotemark{k}$&2&$21.4\pm1.2$&
$2.19\pm0.02$&$0.07^{+.03}_{-.01}$&1.28\\
GeV J0634+1746&Geminga&$0.29\pm0.03$&&$74.3\pm2.2$&
$1.66\pm0.01$&$0.12^{+.04}_{-.02}$&0.32\\
GeV J0835-4512&Vela&$72$&0.5&$148.1\pm3.9$&
$1.69\pm0.01$&$0.17^{+.07}_{-.05}$&0.68\\
GeV J1047-5840&PSR B1046-58&$0.40\pm0.16$&3&$7.2\pm1.4$&
$1.97\pm0.09$&$0.00^{+.18}_{-.00}$&0.56(3) \\
GeV J1059-5218&PSR B1055-52&$0.07\pm0.03$&1.5&$4.2\pm0.9$&
$1.94\pm0.10$&$0.00^{+.25}_{-.00}$&0.47(3)\\
&clump 2\tablenotemark{l}&$0.5$&1.5&$4.2\pm0.9$&
-&-&0.62\\
GeV J1256-0546&3C 279&$10\pm1\tablenotemark{j}$&-&$6.9\pm0.7$&
$1.96\pm0.04$&$1.60^{+1.11}_{-.49}$&0.75\\
GeV J1626-2955&PKS 1622-297&-&-&$4.9\pm0.8$&$2.07\pm0.07$&
$4.56^{+13.62}_{-2.33}$&-\\
GeV J1636+3812&QSO 1633+382&$2.1\pm0.3\tablenotemark{j}$&-&$4.8\pm1.1$&$2.15\pm0.09$&
$0.35^{+.36}_{-.22}$&0.65(2)\\
GeV J1709-4430&PSR B1706-44&$0.95\pm0.30\tablenotemark{m}$&1.8&$19.7\pm1.7$&
$1.86\pm0.04$&$0.16^{+.11}_{-.10}$&0.52(2)\\
\enddata
\tablenotetext{a}{Fluxes are from power law + absorption fits}
\tablenotetext{b}{Distances estimated from nearby young objects for unknown sources \citep{yr97}}
\tablenotetext{c}{Sources with nearby 3EG sources were refit. Otherwise, from
\citet{lm97}}
\tablenotetext{d}{Number in parentheses is error on last digit, if greater than 1}
\tablenotetext{e}{Total non-thermal flux from \citet{s97}}
\tablenotetext{f}{\citet{lhy97}}
\tablenotetext{g}{\citet{s98}}
\tablenotetext{h}{Upper limit on additional power law component with 
spectral index 2.0 after fitting thermal spectrum}
\tablenotetext{i}{Spectral fits from \citet{h96}}
\tablenotetext{j}{\citet{k98}}
\tablenotetext{k}{\citet{z90}}
\tablenotetext{l}{Values are for the bright, extended source,
(clump 2) from \citet{sh97}}
\tablenotetext{m}{\citet{f98}}
\end{deluxetable}

\clearpage

\begin{deluxetable}{lccccccccl}
\tablecaption{X-Ray Sources in New Fields \label{NEWfit}}
\tablewidth{0pt}
\tabletypesize{\scriptsize}
\tablehead{
\colhead{$\gamma-$ray source}&\colhead{X-ray source}&\colhead{R.A.(2000)}&
\colhead{Dec.(2000)}&
\colhead{nH}&\colhead{$\Gamma$\tablenotemark{a}}&
\colhead{$F_{2-10keV}$}&\colhead{$N(>F)$\tablenotemark{b}}&
\colhead{ID} & Notes\\		
&&h m s& ${^{\circ}} {^{\prime}} {^{\prime \prime}}$&$10^{22}{\rm cm}^{-2}$&&$10^{-12}$ ergs ${\rm cm}^{-2}{\rm s}^{-1}$&&&
}
\startdata
GeV J1025-5809&AX J1025.6-5757&10 25 35.8&-57 56 47&
$0.58^{+2.95}_{-.58}$&$1.29^{+1.70}_{-1.01}$&$0.34\pm0.07$&1.7&src1 &\\
&AX J1025.9-5749&10 25 56.6&-57 48 42& $1.19^{+.32}_{-.27}$&
$2.90^{+.38}_{-.34}$&$2.59\pm0.09$&0.3&src2, 1E1024.0-5732& d,o\\
GeV J1417-6100&AX J1420.1-6049&14 20 07.8&-60 48 56&
$1.79^{+.72}_{-.49}$&$1.36^{+.26}_{-.36}$&$4.82\pm0.17$&0.2&K3, obs2&e\\
&AX J1418.7-6058&14 18 39.8&-60 58 03&$1.81^{+.33}_{-.30}$&
$1.86^{+.20}_{-.20}$&$7.33\pm0.17$&0.1&Rabbit, obs2&e\\
&AX J1418.2-6047&14 18 12.8&-60 46 55&$2.79^{2.37}_{-2.00}$&
$1.26^{+.95}_{-1.04}$&$0.98\pm0.11$&0.7&src2, obs1&c\\
&AX J1418.6-6045&14 18 37.0&-60 45 12&$3.27^{+3.83}_{-2.33}$&
$1.14^{+1.11}_{-1.03}$&$0.97\pm0.11$&0.7&src3, obs1&c\\
GeV J1732-3130&AX J1733.9-3112&17 33 52.3&-31 12 25&
$1$&$2$&$0.46\pm0.16$&2.1&src1&d\\
&AX J1732.2-3044&17 32 09.9&-30 43 49&$0.43^{+1.44}_{-.43}$&
$2.08^{+1.88}_{-1.00}$&$1.35\pm0.23$&0.8&src2&d\\
&AX J1731.7-3115&17 31 41.2&-31 15 28&$0.68^{+1.22}_{-.68}$&
$2.69^{+2.39}_{-1.22}$&$0.45\pm0.07$&2.1&src3&c \\
&AX J1732-3115&17 32 01&-31 14 38&$0.51^{+.66}_{-.44}$&$1.92^{+.62}_{-.53}$&
$2.40\pm0.20$&0.5&src4,diffuse &f\\
GeV J1809-2327&AX J1809.8-2333&18 09 47.1&-23 33 17&$1.67^{+.35}_{-.31}$&
$2.23^{+.23}_{-.22}$&$8.61\pm0.21$&0.02&Nebula&e\\
&AX J1809.8-2332&18 09 48.6&-23 32 09&$1.77^{+.29}_{-.26}$&
$2.09^{+.18}_{-.17}$&$3.65\pm0.08$&(0.05)&sub-peak 1&c\\
&AX J1809.8-2339&18 09 49.8&-23 38 51&$0.51^{+.31}_{-.24}$&$2.63^{+.46}_{-.36}$
&$0.67\pm0.03$&(0.2)&sub-peak 2&c,o\\
&AX J1810.0-2340&18 09 57.2&-23 39 47&$0.61^{+.39}_{-.32}$&$2.61^{+.56}_{-.47}$&
$0.60\pm0.03$&(0.3)&sub-peak 3&c,o\\
&AX J1809.9-2336&18 09 55.7&-23 35 51&$1.83^{+.84}_{-.57}$&
$2.21^{+.57}_{-.41}$&$1.01\pm0.06$&(0.2)&sub-peak 4&c\\
&AX J1810.6-2349&18 10 39.5&-23 48 42&$2.70^{+1.16}_{-1.74}$&
$0.85^{+.68}_{-.69}$&$2.48\pm0.20$&&src2 AGN?&d,n\\
GeV J1825-1310&AX J1826.1-1300&18 26 04.9&-12 59 48&$1.53^{+.36}_{-.32}$&
$2.17^{+.25}_{-.24}$&$8.27\pm0.24$&0.2&Nebula&e\\
&AX J1826.1-1257&18 26 08.2&-12 56 46&$1.61^{+1.08}_{-.76}$&$2.02^{+.78}_{-.61}$&
$1.43\pm0.11$&(0.7)&sub-peak 1&c \\
&AX J1826.0-1303&18 26 01.2&-13 02 54&$0.77^{+.61}_{-.39}$&$1.92^{+.62}_{-.47}$&
$1.25\pm0.08$&(0.8)&sub-peak 2 &c\\
&AX J1824.5-1310&18 24 32&-13 09 59&$5.22^{+3.64}_{-2.28}$&
$6.96^{+3.04}_{-2.33}$&$1.09\pm0.19$&0.9&src3, G18.1-0.2&e,o\\
GeV J1837-0610&AX J1837.5-0610&18 37 32.5&-06 09 49&$0.58^{+1.17}_{-.58}$&
$0.48^{+.55}_{-.57}$&$1.63\pm0.15$&0.1&&c\\
GeV J1907+0557&AX J1907.4+0549&19 07 21.3&+05 49 14&$5.09^{+5.89}_{-2.48}$&2&
$0.76\pm0.17$&1.0&&e? \\
GeV J2020+3658&AX J2021.6+3656&20 21 33.2&+36 55 36&$0.97^{+.35}_{-.26}$&
$1.95^{+.29}_{-.22}$&$3.13\pm0.09$&0.1&src1,WR141&c\\
&AX J2021.1+3651&20 21 07.8&+36 51 19&$0.50^{+.25}_{-.25}$&
$1.73^{+.26}_{-.28}$&$3.83\pm0.13$&0.1&src2&c\\
GeV J2026+4124&AX J2027.6+4116&20 27 33.8&+41 16 12&1&2&$1.00\pm0.69$&1.2&&g,d\\
GeV J2035+4214&AX J2036.0+4218&20 35 57.6&+42 17 38&$2.95^{+2.21}_{-1.43}$&
$1.41^{+.73}_{-.57}$&$1.39\pm0.10$&0.8&src1& \\
&AX J2035.4+4222&20 35 24.3&+42 22 04&$0.98^{+.71}_{-.45}$&$2.02^{+.63}_{-.44}$&
$0.88\pm0.06$&1.2&src2&\\
&AX J2035.9+4229&20 35 55.2&+42 29 09&$0.84^{.52}_{-.37}$&$2.44^{+.70}_{-.50}$&
$1.14\pm0.07$&0.9&src3&e \\
\enddata
\tablenotetext{a}{Photon spectral index.}
\tablenotetext{b}{Number of sources with flux greater than the source expected
in 95\% error contour, from log N-log S distribution of Sugizaki (1999).} 
\tablenotetext{c}{Source spectrum (and hence flux measurement) possibly
contains photons from nearby source or diffuse emission in addition to 
any point source contribution.}
\tablenotetext{d}{Source is near edge of detector. May have large systematic 
positional error.}
\tablenotetext{e}{Source appears extended.}
\tablenotetext{f}{Possibly scattered emission from X1724-308}
\tablenotetext{g}{Flux highly uncertain due to scattered emission from
Cyg X-3}
\tablenotetext{n}{Source outside 99\% $\gamma-$ray contour. Unlikely to
be counterpart.}
\tablenotetext{o}{Source outside 95\% $\gamma-$ray contour.}

\end{deluxetable}

\end{document}